\def   \ni {\noindent}

\def   \ssk {\vskip  5truept}

\def   \bsk {\vskip 14truept}
 
\def   \newpage {\vfill\eject}
\def   \newline {\hfil\break}

\def\loe{\lower 0.6ex\hbox{${}\stackrel{<}{\sim}{}$}}
\def\goe{\lower 0.6ex\hbox{${}\stackrel{>}{\sim}{}$}}

\documentstyle[epsfig]{article}
\begin{document}

\hsize 5truein
\vsize 8truein
\font\abstract=cmr8
\font\keywords=cmr8
\font\caption=cmr8
\font\references=cmr8
\font\text=cmr10
\font\affiliation=cmssi10
\font\author=cmss10
\font\mc=cmss8
\font\title=cmssbx10 scaled\magstep2
\font\alcit=cmti7 scaled\magstephalf
\font\alcin=cmr6 
\font\ita=cmti8
\font\mma=cmr8
\def\ref{\par\noindent\hangindent 15pt}
\null


\vspace*{-0.5cm}
\fnsymbol{footnote}
\title{\ni GAMMA-RAY BURSTS: THE FOUR CRISES  $^{^{\Huge(\star)}}$}

\bsk \bsk
\author{\ni M.~Tavani $^{1,2}$}
\bsk
\affiliation{
\noindent
(1) Istituto Fisica Cosmica, CNR, via Bassini 15, Milano I-20133 (Italy)\\
(2) Columbia Astrophysics Lab., Columbia University, New York, NY 10027 (USA)} 

\bsk
\baselineskip = 12pt

\abstract{ABSTRACT \ni
We discuss some open problems concerning the origin and the
emission mechanism of gamma-ray bursts (GRBs) in light of
recent developments.
If GRBs originate at extragalactic distances, we are facing four
crises:
{\it (1) an energy crisis}, models have to account for more than $10^{53}$~ergs
of energy emitted in the gamma-ray band;
{\it (2) a spectral crisis}, emission models have to account for the
 surprising
`smoothness' of GRB broad-band spectra, with no indication of the predicted
spectral `distorsions' caused by inverse Compton scattering in large
radiation energy density media, and no evidence for beaming;
{\it (3) an afterglow crisis}, relativistic shock
 models have to explain the complexity
of the afterglow behavior, the longevity of optical transients detectable 
up to six months after the burst,
the erratic  behavior of the radio emission,
the lack of evidence for substantial beaming as indicated by
recent searches for GRB afterglows in the X-ray band;
{\it (4) a population crisis}, from data clearly indicating that {\it only}
hard and long GRBs  show a strong deviation from an Euclidean brightness
distribution, just the opposite of what expected from extragalactic models
without substantial cosmological evolution.
All previously proposed cosmological models are challenged:
in particular, the neutron star-neutron  star coalescence model
most likely will not survive the resolution of the problems
raised by points (1) and (4).
}                                                    


\bsk
\baselineskip = 12pt


\text{\ni 1. INTRODUCTION
\ssk
\ni     

Recent observations suggest that gamma-ray bursts (GRBs)
might  originate at extragalactic distances (for a review, 
and references concerning observations, see Hurley 1998).
However, a satisfactory resolution of the question regarding the ultimate
source of the GRB phenomenon still escapes us.
As usually happens in rapidly evolving fields, observational puzzles
  outnumber
theoretical explanations. Observations  
provide a very complex scenario. From the handful of
 optical transients (OTs) detected  within the error boxes of 
X-ray afterglow sources, GRBs appear to originate at large 
distances ($z \goe 1-2$). Is this emerging picture proved ?
Does the extragalactic origin apply to all GRBs ?
In this paper we briefly  discuss four  problematic issues 
requiring drastic resolutions  within the framework of
extragalactic GRB models. The page limit does not allow
a discussion of the qualitative successes of the
relativistic fireball model, or a treatment of alternative models.
We will assume, for the sake of the argument, that all proposed
associations of OTs and extragalactic hosts
with GRBs are correct.

\noindent
--------------------------------------------------------------------------\\
($ ^{\star}$)
 Adapted from a paper presented on September 18, 1998
in Taormina, to be published in the Proceedings
of the 3rd INTEGRAL Workshop {\it The Extreme Universe}.

\newpage

\bsk
\ni 2. THE FOUR CRISES
\ssk
\ni
\noindent
{\it (1) The energy crisis}

The implied enormous energies for isotropic emission
 in the  hard X-ray/soft gamma-ray band
($>  10^{53} $~ergs as deduced  directly for GRB~971214 or indirectly
for GRB~970228 by the absence of spectroscopic or photometric redshift 
of the OT's host, implying $1 \loe z \loe 2$) are problematic for any model.
The neutron star-neutron star (NS-NS) models  proposed a few years ago
were  characterized by a few percent of the energy 
($\sim 10$\% of the total mass-energy)
 available for high-energy emission, i.e.,  $10^{51}$~ergs
(e.g., Paczynski 1990, Meszaros \& Rees 1993).
Clearly  NS-NS models require substantial  beaming to explain new
GRB-OT associations, or a very efficient mechanism of energy conversion
for which we do not have any other indication of its existence. 
NS-NS models also suffer from the relative closeness of OT's to
their host galaxies, contradicting  calculations of the
 NS-NS coalescing binary evolution.

 We are left with  three main alternatives:
(1) {\it  hypernovae }
(a concept introduced after the detection of GRB~970508, see Paczynski 1998);
(2) a special class of {\it failed supernovae}
 (e.g., MacFadyen \& Woosley 1998);
and (3)
{\it relatively massive black holes}  with their rotational and gravitational
energy transformed  by very efficient electromagnetic/MHD  effects (e.g., 
Meszaros, Rees \& Wijers 1998)
or quantum effects (e.g., Preparata et al. 1998)
into relativistic particle outflows.
%
Even before addressing  the theoretical details of these models, 
one crucial aspect is clear:
all models of this type require relativistic beaming to explain GRB
light curves.
 However,
 there is no evidence 
today that the GRB prompt  emission is beamed.
A recent analysis
of GRB lightcurves simulated by relativistically beamed emitting fronts
fails to reconcile data with theoretical expectations (Fenimore et al. 1998). 

\ssk
\ni
{\it (2) The spectral crisis}

Broad-band GRB spectra are remarkably smooth.
This property
 is in contrast with lightcurves, that are often very complex.
The `smoothness' of GRB spectra applies to both time-averaged spectra
(Tavani 1996a,b) and also to time-resolved spectra, as indicated in BATSE
data (e.g., Band 1996), and in BSAX data covering the $\sim 3-300$~keV
 energy range  (e.g., Piro et al. 1998, Frontera et al. 1998, 1999).
Efficient particle acceleration and synchrotron emission 
can naturally reproduce  the majority of available time-resolved
spectra (Tavani 1996b, 1997). What is now called the Shock
Synchrotron Model (SSM) qualitatively agrees with expectations from
a particular class of relativistic fireball models (e.g., Meszaros et al.
1994, hereafter MRP94). (Low-energy suppression  applies
to a minority of time-resolved spectra, possibly indicating a
violation of optically thin conditions during the initial
phase of some GRB pulses, see Tavani 1998b; 
for a different approach, Liang 1997).

However, the agreement  of data with an optically thin SSM with no
spectral distorsions 
is surprising. Strong deviations from the simplest SSM spectral
shape are expected if reverse and forward shock contributions to
the prompt emission operate simultaneously or at different times. 
Strong modifications by inverse Compton scattering are expected in
the majority of relativistic shock front for cosmological GRB models
(e.g., MRP94, Pilla \& Loeb  1998). None are observed.
`Piston models' (e.g., MRP94) might explain the absence of multicomponent
spectra, even though it is not clear why mixing of the reverse and
forward shock fronts would apply to all GRBs.
In any case, the absence of IC distorsions strongly constrains
extragalactic  models, and no satisfactory explanation
 has yet been proposed.

\ssk
\ni
{\it (3) The afterglow crisis}

The remarkable discovery of detectable
 X-ray afterglows lasting hours/days/weeks clearly demonstrates
that the GRB phenomenon can dissipate a very large fraction of the
total energy at very late times (Costa et al., 1997, 1998).
This is in contrast with pre-afterglow models of GRB
fireball dissipation.
Decelerating relativistic shock fronts can generate radio/optical/X-ray
emission only for very efficient particle acceleration acting during
the initial impulsive phase and most likely even at later times.
 It is not clear whether the broad-band
aftergloe emission can be understood in
terms of decelerating relativistic fronts with
simple energized particle injections
(e.g., Sari, Piran \& Narayan 1998).
Clearly, the efficient and probably 
 prolonged particle acceleration phases of GRB afterglows require a 
more detailed theoretical analysis.
In addition, no evidence exists for an enhanced GRB rate
expected in models of afterglow X-ray emission dominated 
by strong beaming effects.
No GRB afterglows more than expected were found on timescales of hours/days 
in the ROSAT All Sky Survey database (Greiner 1998). 
It is also interesting that  beaming does not
affect the rate of GRB prompt X-ray emission as deduced by the null results
of a search in  X-ray  all-sky survey archival data
(Grindlay 1999).

\ssk
\ni
{\it (4) The population crisis}

We know relatively little about the short GRBs with duration
$\tau_b < 2.5$~s. They are usually quite hard, and may
be thought as more `elementary' than the usually  multi-peaked 
bursts of longer duration. It is interesting that the brightness
distribution (logN-logP)
for this subclass of bursts shows little deviation from
an Euclidean three-dimensional distribution. They can be classified
as Short/Hard (S/H) bursts, typically with a BATSE's
hardness ratio $H_{32} \goe 3$.
 No counterpart for these bursts has yet 
been detected, and in principle they can be anywhere (see also Hurley 1998).

Long duration events ($\tau_b >  2.5$~s) constitute the
majority of detected GRBs, and if derived from a source population at
extragalactic distances (e.g., star forming galaxies)
with no substantial cosmological evolution of  burst properties
(spectrum, intensity), their brightness distribution is expected to
 be strongly non-Euclidean especially  for the 
Long/Soft (L/S) subclass (i.e., for  long GRBs with hardness ratio
in the ideal range to be detected by BATSE, $H_{32} \loe 3$).

This expectation is in direct contradiction with the data 
(Tavani 1998a, hereafter T98a). It is surprising for extragalactic models with
no evolution  that  {\it only} 
Long/Hard (L/H) bursts (with $H_{32} > 3$) have a
strongly non-Euclidean brightness distribution. Remarkably, the
brightness distribution of L/S bursts is very similar to that for 
S/H bursts. Fig.~1 shows the brightness distributions  of these
subclasses as derived from
the {\it average properties} of GRBs from the  4-th BATSE Catalogue.

This `population problem' requires  a drastic  cure.
Two possibilities arise:
{\it (i)} source evolution at large redshifts has to produce GRBs on
the average much harder (on the borderline for being detected 
efficiently by BATSE) and more intense than those at smaller redshifts;
or 
{\it (ii)} BATSE's  selection effects  have to
play a major role in producing an
{\it apparent} non-Euclidean and inhomogeneous
 distribution only for  L/H bursts
(these bursts typically show emission outside the ideal energy range 
for BATSE's  triggers,
50--300~keV, see T98a). 
 The fact that the brightness distributions of spectral
(hard/soft) subclasses derived from the  {\it individual properties of
GRB pulses} (Pendleton et al., 1997)
 are similar to those derived from the average properties (T98a)
 favors the latter explanation.
More work is needed to clarify this important issue.



\bsk
\baselineskip = 12pt


{\references \ni REFERENCES
\ssk

\ref Costa, E., et al., 1997, Nature, 387, 783.

\ref Costa, E., et al., 1998, paper presented at the Conference
{\small \it GRBs in the Afterglow Era}, Rome 3-6 November, 1998. Proceedings
to be published in the A\&A Suppl. Series.

\ref Fenimore, E.E., et al., 1998, submitted to ApJ 
(astro-ph/9802200). 

\ref Frontera, F., et al., 1998, paper presented at the Conference
{\small \it GRBs in the Afterglow Era}, Rome 3-6 November, 1998. Proceedings
to be published in the A\&A Suppl. Series.

\ref Frontera, F., et al., 1999, in preparation.

\ref Greiner, J, 1998, paper presented at the Conference 
{\small \it GRBs in the Afterglow Era}, Rome 3-6 November, 1998. Proceedings
to be published in the A\&A Suppl. Series.

\ref Grindlay, J., 1999, ApJ, in press (astro-ph/9808242).

\ref Hurley, K., 1998, these Proceedings.

\ref Liang, E.P., 1997, ApJ, 491, L15.

\ref{MacFadyen, A. \& Woosley, S.E., 1998,
submitted to ApJ (astro-ph/9810274).

\ref M\'esz\'aros, P., Rees, M.J., 1993, ApJ, 405, 278.

\ref M\'esz\'aros, P., Rees, M.J. \& Papathanassiou, H., 1994, ApJ, 432, 181.
 
\ref M\'esz\'aros, P., Rees, M.J. \& Wijers, R.A.M.J., 1998, (astro-ph/9808106).

\ref Paczy\'nski, B. 1990, ApJ, 363, 218.

\ref Paczy\'nski, B. 1998, ApJ, 494, L45.

\ref Pendleton, G.N., et al., 1997, ApJ, 489, 175.

\ref Pilla, R.P. \& Loeb, A., 1998, ApJ, 494, L167.

\ref Piro, L., et al., 1998, A\&A, 331, L41.

\ref Preparata, G., Ruffini, R. \& Xue, S-S., 1998,
A\&A, in press (astro-ph/9810182).

\ref Sari, R., Piran, T. \& Narayan, R., 1998, ApJ, 497, L17.

\ref Tavani, M., 1996a, PRL, 76, 3478.

\ref Tavani, M., 1996b, ApJ, 466, 768.

\ref Tavani, M., 1997, ApJ, 480, 351.

\ref Tavani, M., 1998a, ApJ, 497, 21.

\ref Tavani, M., 1998b, in preparation.
}

\newpage

\begin{figure}[!ht]
\centerline{\psfig{file=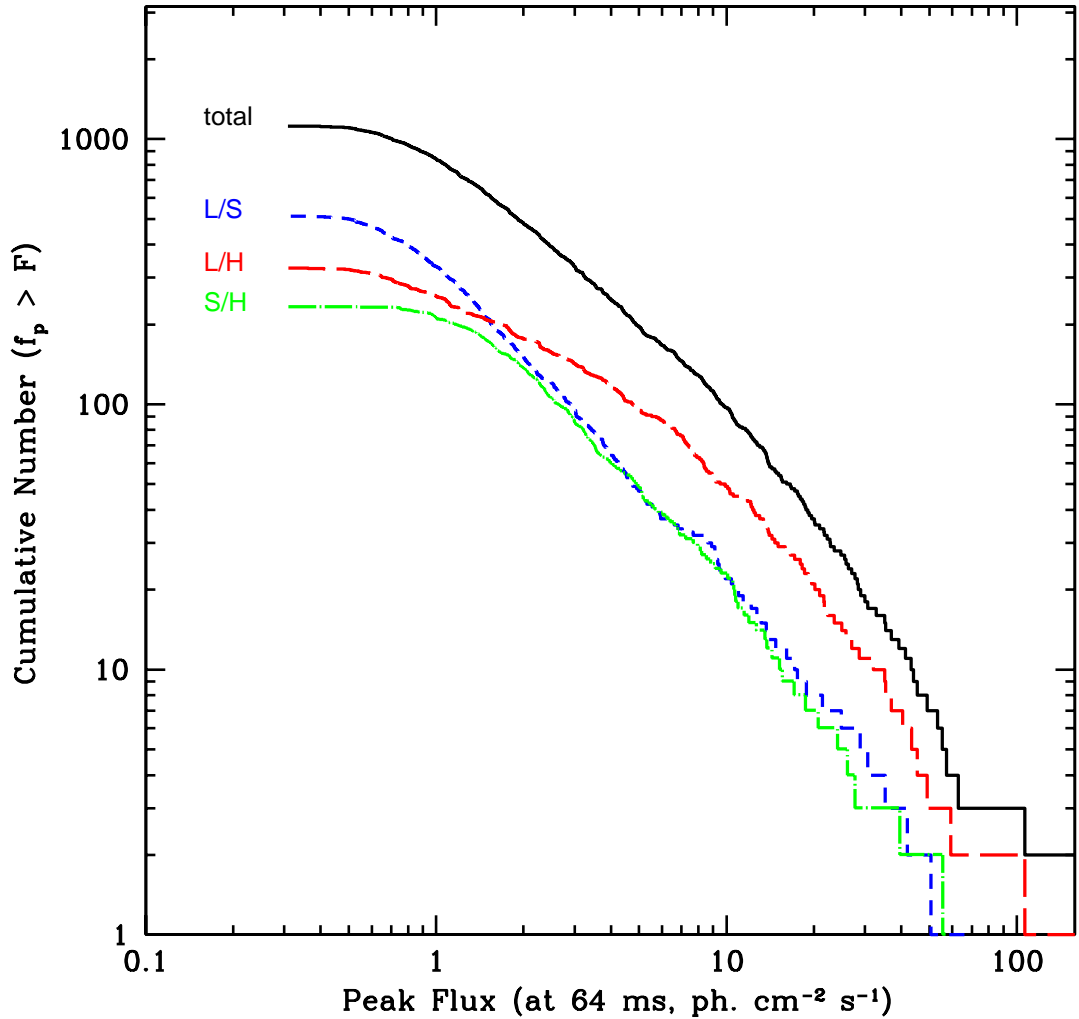, width=15cm}}
\caption{FIGURE 1. BATSE's brightness distributions  (log$N$-log$P$,
with $N$ the cumulative number of bursts above a given peak photon countrate
$P$ in the energy range 50-300~keV and for the 64~ms time resolution) 
for different GRB sub-classes (from Tavani 1998a).
Distributions for the three main GRB sub-classes are shown:\\
(S/H) GRBs with $T_{90} < 2.5$ sec, and $H^e_{32} > 3$ (short/hard bursts);\\
(L/H) GRBs with $T_{90} > 2.5$ sec, and $H^e_{32} > 3$ (long/hard bursts);\\ 
(L/S) GRBs with $T_{90} > 2.5$ sec, and $H^e_{32} < 3$ (long/soft bursts).
}
\end{figure}

\end{document}